# Hydrophobic interaction determines docking affinity of SARS-CoV-2 variants with antibodies

Jiacheng Li[a,1], Chengyu Hou[b,1], Menghao Wang[a,1], Chencheng Liao[b,1], Shuai Guo[a], Liping Shi[a], Xiaoliang Ma[a], Hongchi Zhang[a], Shenda Jiang[a], Bing Zheng[c], Lin Ye[d], Lin Yang[a,d,*], Xiaodong He[a,e,*]

[a] *National Key Laboratory of Science and Technology on Advanced Composites in Special Environments, Center for Composite Materials and Structures, Harbin Institute of Technology, Harbin 150080, China*

[b] *School of Electronics and Information Engineering, Harbin Institute of Technology, Harbin 150080, China*

[c] *Key Laboratory of Functional Inorganic Material Chemistry (Ministry of Education) and School of Chemistry and Materials Science, Heilongjiang University, Harbin 150001, P. R. China.*

[d] *School of Aerospace, Mechanical and Mechatronic Engineering, The University of Sydney, NSW 2006, Australia*

[e] *Shenzhen STRONG Advanced Materials Research Institute Co., Ltd, Shenzhen 518035, P. R. China.*

**Abstract**

Preliminary epidemiologic, phylogenetic and clinical findings suggest that several novel severe acute respiratory syndrome coronavirus 2 (SARS-CoV-2) variants have increased transmissibility and decreased efficacy of several existing vaccines. Four mutations in the receptor-binding domain (RBD) of the spike protein that are reported to contribute to increased transmission. Understanding physical mechanism responsible for the affinity enhancement between the SARS-CoV-2 variants and ACE2 is the "urgent challenge" for developing blockers, vaccines and therapeutic antibodies against the coronavirus disease 2019 (COVID-19) pandemic. Based on a hydrophobic-interaction-based protein docking mechanism, this study reveals that the mutation N501Y obviously increased the hydrophobic attraction and decrease hydrophilic repulsion between the RBD and ACE2 that most likely caused the transmissibility increment of the variants. By analyzing the mutation-induced hydrophobic surface changes in the attraction and repulsion at the binding site of the complexes of the SARS-CoV-2 variants and antibodies, we found out that all the mutations of N501Y, E484K, K417N and L452R can selectively decrease or increase their binding affinity with some antibodies.

**Introduction**

*Corresponding author. E-mail address: linyang@hit.edu.cn (Lin Yang) [1]These authors contributed equally to this work.

The emergence of several variants of the novel severe acute respiratory syndrome coronavirus 2 (SARS-CoV-2) undoubtedly brings new challenges to the development of vaccines and therapeutic antibodies. A key question is whether existing COVID-19 vaccines can protect against infection or disease from these new SARS-CoV-2 variants. Due to the emergence of a new SARS-CoV-2 variant 501Y.V1 (B.1.1.7) in the United Kingdom, there was an unexpected increase in reported COVID-19 cases (*1*). In South Africa, a high transmission in the context of high population immunity are most likely caused by the emergence of another SARS-CoV-2 variant 501Y.V2 (B.1.351)(*2*). Both the variants have a mutation (N501Y) in the receptor binding domain of the spike protein, which is reported to help increase transmission, and the increased transmission range is estimated to be between 40% and 70% (*2*). The 6501Y.V2 variant also has two mutations in the spike protein (E484K and K417N) that give the antibody a potential immune escape opportunity(*3*). The preliminary clinical trial results of the ChAdOx1 nCoV-19 vaccine showed 74% effectiveness in the United Kingdom, but only 22% in South Africa(*4*). While another COVID-19 vaccine NVX-CoV2373 showed 89% effectiveness in the United Kingdom, but in South Africa, only 49%(*5*).

All SARS-CoV-2 variants enter human cells by protein-protein docking to human angiotensin converting enzyme 2 (ACE2) on the host cell membrane via CoV spike (S) glycoproteins. Molecular structures of the S protein of SARS-CoV-2 have been observed at high resolution by using cryo-electron microscopy (cryo-EM) (*6-8*). The complex structures of ACE2 bound to the SARS-CoV-2 S have also been experimentally determined (*9-12*). Surprisingly, a change from asparagine (N) to tyrosine (Y) in amino-acid position 501 of the RBD of SARS-CoV-2 S can obviously increase transmission, with estimates ranging between 40% and 70% for increased transmission.

The region of the protein responsible for binding another molecule is known as the docking site (also sometimes called binding site) and is often a depression on the molecular surface. The key to SARS-CoV-2 infection is that the S protein can specifically bind to the ACE2 in a strong affinity manner. Specific binding of SARS-CoV-2 S and ACE2 forms the joint structure between the coronavirus and the host cell that enable the coronavirus enter the host cell (*13*). In natural intracellular environment and extracellular medium, protein-protein docking are usually the contacts of high specificity established between two or more specific protein molecules, and erroneous protein-protein docking rarely occurs (*14*). The chief characteristic of proteins that allows their diverse set of functions is their ability to dock with other proteins specifically and tightly. Protein-protein docking is therefore considered one of the miracles of nature. The classic problem of protein-protein docking is the question of how a protein find its partner in its natural environment (*15*). In particular, whether hydrogen bonds formation regulate protein-protein docking remains a long-standing problem with poorly defined mechanisms (*16-21*). It is worth noting that hydrogen bonds formation is not a long-range physical force. Considering that only several hydrogen bonds between SARS-CoV-2 RBD and ACE2 can be identified in the complex (*9-12, 22*), the docking between the coronavirus and the host cell are most likely not dominated by hydrogen bond pairing between them.

A recent theoretical study revolutionize the understanding of docking affinity of SARS-CoV-2 S with ACE2 and the physical mechanism for protein docking. Traditional theories hold that, protein-protein docking is guided by a variety of physical forces as follows: (i) hydrophobic effect, (ii) electrostatic forces, (iii) van der Waals forces, (iv) hydrogen bonding, (v) ionic bonding, (vi) entropy. While the new theoretical study emphasizes that only hydrophobic effect can play a decisive role for protein docking (*16, 23*). In extracellular medium, hydrogen-bond competing is always present with water. Because, water molecules should be able to saturate the hydrogen bond formations of the hydrophilic surface of proteins before the protein docking (*24-26*), due to water molecules having very strong polarity (*26-28*). The interaction of protein surface with the surrounding water is often referred to as protein hydration shell (also sometimes called hydration layer) and is fundamental to structural stability of protein, because non-aqueous solvents in general denature proteins (*29*). The hydration layer around a protein has been found to have dynamics distinct from the bulk water to a distance of 1 nm and water molecules slow down greatly when they encounter a protein(*24*). Thus, hydrophilic side chains of proteins are normally hydrogen bonded with surrounding water molecules in aqueous environments, thereby preventing the surface hydrophilic side-chains of proteins from randomly hydrogen-bonding together (*24, 25*) (*26*). This is the reason why proteins usually do not aggregate and crystallize in unsaturated aqueous solutions(*30*). Thus, this problem may lie in our lack of understanding of how surface hydrophilic side-chains of a protein at the docking site can get rid of their hydrogen-bonded water molecules, then hydrogen-bond with other hydrophilic groups or approach to hydrophobic groups of a partner protein at early steps of the docking, given the lack of awareness of the importance of the shielding effect of water.

The hydrophobic interaction between hydrophobic surficial areas of two proteins at the docking site can reintroduce entropy to the system via the breaking of their water cages which frees the ordered water molecules(*31*). Thus, the hydrophobic collapse between hydrophobic surface areas of a protein and hydrophobic surface areas of another protein most likely triggers enthalpy-entropy compensation for the docking of the two proteins, enabling the hydrophilic groups at the docking site to discard hydrogen-bonded water molecules, and then hydrogen-bond with each other between the proteins or approach to hydrophobic groups of the partner protein. This means that the hydrophobic interaction among proteins should be responsible for the protein docking. The hydrophobic interaction between the SARS-CoV-2 S and ACE2 protein is found to be significantly greater than that between SARS-CoV S and ACE2 (*32*). This explains why the affinity of SARS-CoV-2 RBD and ACE2 far exceeds that of SARS-CoV RBD and ACE2.

**Results**

Binding ability of a protein is mediated by the tertiary structure of the protein, which defines the docking site, and by the chemical properties of the surrounding amino acids' side chains (*33*). The hydrophobicity of the protein surface is the main factor that stabilizes the protein-protein binding, thus hydrophobic interaction among proteins play an important role in determining the protein-protein binding affinity (*23, 34, 35*). Hydrophilic side-chains are not completely hydrophilic. The hydrophilicity of hydrophilic side-chains is normally expressed by C=O or N-H2 groups at their ends, and the other portions of hydrophilic side-chains are hydrophobic, because the molecular structures of these portions are basically alkyl and benzene ring structures. It means that there are a large number of water molecules surround the hydrophobic surface areas of the RBD rather than hydrogen bonded with the RBD. The characteristic of these water molecules surrounding the hydrophobic surface areas is that their hydrogen bonding network is more ordered than free liquid water molecules, that is, their entropy is lower. At the docking site, a hydrophobic surface area of a protein facing to another hydrophobic surface area of the other protein play a role of increasing the binding affinity between of the two proteins. Whereas, a hydrophobic surface area of a protein facing to another hydrophilic surface area of the other protein play a role of decreasing the binding affinity between of the two proteins due to the repulsion. Thus, mutation of one amino acid residue at the docking site may can obviously change the hydrophobic interactive surface areas of the SARS-CoV-2 RBD at the docking site and may significantly change the hydrophobic interaction between of the SARS-CoV-2 S and ACE2, thereby greatly change the affinity between them.

For example, the original hydrophilic asparagine (N) in amino-acid position 501 of the RBD was facing to a hydrophobic surface area one the ACE2 at the docking site. The change from asparagine (N) to tyrosine (Y) in amino-acid position 501 increase hydrophobic attraction area and decrease hydrophilic repulsion area between the two proteins, see Fig.1. Therefore, a change from asparagine (N) to tyrosine (Y) in amino-acid position 501 of the RBD can change the original repulsing relationship between the RBD and ACE2 to an attractive relation between them in amino-acid position 501. By analyzing the mutation-induced surface hydrophobic or hydrophilic area changes at the binding site of the complex of the SARS-CoV-2 variants and ACE2, we found out that only the mutations N501Y obviously increase the binding affinity between the RBD to the ACE2, and the mutation E484K and L452R slightly increase the binding affinity between the RBD to the ACE2, see Fig.1 and Table 1. It is worth noting that the change from Leucine (L) to arginine (R) in amino-acid position 452 due to the mutation L452R introduce a new arginine-glutamic acid (R-E) electrostatic attraction between the RBD and the ACE2 that may cause increment of the binding affinity.

Use same method, we can evaluate the impact of these mutations in the RBD on protection capability of the existing antibodies. By analyzing the mutation-induced surface hydrophobic or hydrophilic area changes at the binding site of the complex of the SARS-CoV-2 variants and 9 existing antibodies, we found out that the mutation N501Y can selectively decrease the binding affinity with the antibodies CV30, and CC12.1, see the From 1. The mutation E484K can selectively decrease the binding affinity with the antibodies B38. The mutation K417N can decrease the binding affinity with the antibodies B38, CC12.3 and COVA2-39. The mutation E484K can selectively decrease the binding affinity with the antibodies B38. The mutation L452R can selectively decrease the binding affinity with the antibodies CB6, SR4, CV30, CC12.1, CC12.3, P2B-2F6, MR17-K99Y, and B38. Taken together, only the antibody CB6 and CV30 was barely able to protect against infection from all these new variants without obviously decreasing the protective efficacy.

**Conclusion**

The higher transmissibility of the variants 501Y.V1 and 501Y.V2 are most likely resourced from increment of hydrophobic attraction and decrement of hydrophilic repulsion between the RBD and ACE2, due to the change from asparagine (N) to tyrosine (Y) in amino-acid position 501 of the RBD. Thus, the only one amino acid residue mutation not only increase the degree of hydrophobic paring between SARS-CoV-2 RBD and ACE2 but also decrease the degree of hydrophobic-hydrophilic paring between them, that greatly enhance the affinity of SARS-CoV-2 and ACE2 receptor and obviously increase the infectiousness. By analyzing the mutation-induced area changes in the attraction and repulsion at the binding site of the complex of the SARS-CoV-2 variants and antibodies, we found out that all the mutations of N501Y, E484K, K417N and L452R can selectively decrease or increase the binding affinity of some antibodies, cause only a small minority of the existing antibodies can well protect against infection or disease from all these SARS-CoV-2 variants as expected. Sufficient hydrophobic interaction at the docking site between proteins should be regarded as the most important driving force for the protein docking, providing enthalpy-entropy compensation at the docking site enable the hydrophilic residues in this region get rid of the hydrogen-bonded water molecules, and promote hydrogen bonding and electrostatic attraction among these hydrophilic side-chains at the binding site. Docking affinity between virus variants and the receptor must be sensitive to changes in the distribution of hydrophobic and hydrophilic surface areas in the PDB.

## Materials and Methods

### Protein structures

In this study, many experimentally determined native structures of proteins are used to study the mechanism triggering docking of SARS-CoV-2 variants to ACE2 and the antibodies. All the three-dimensional (3D) structure data of protein molecules are resourced from the PDB database, including the experimentally determined the RBD of SARS-CoV-2 S, ACE2, antibodies and their complexes, et al. IDs of these proteins according to PDB database are marked in all the figures. In order to show the distribution of hydrophobic areas on the surface of the SARS-CoV-2 RBD, ACE2, antibodies and their complexes at the binding sites in these figures, we used the structural biology visualization software PyMOL to display the protein hydrophobic surface areas.

### Calculation of hydrophobic or hydrophilic surface area

Affinity of RBD and ACE2 and antibodies can be characterized by calculating the size of the hydrophobic contact area in the complex structures. We used molecular 3D structure display software PyMOL to draw the hydrophobic and hydrophilic surface areas at the docking sites. We calculated the hydrophobic attraction surface areas and hydrophobic repulsion surface areas in the positions of mutations involved in the hydrophobic interaction among the RBD, ACE2 and antibodies in this study.

### Acknowledgements

Lin Yang is indebted to Daniel Wagner from the Weizmann Institute of Science and Liyong Tong from the University of Sydney for their support and guidance. Lin Yang is grateful for his research experience in the Weizmann Institute of Science for inspiration. The authors acknowledge the financial support from the National Natural Science Foundation of China (Grant 21601054), Shenzhen Science and Technology Program (Grant No. KQTD2016112814303055), Science Foundation of the National Key Laboratory of Science and Technology on Advanced Composites in Special Environments, the Fundamental Research Funds for the Central Universities of China and the University Nursing Program for Young Scholars with Creative Talents in Heilongjiang Province of China (Grants UNPYSCT-2017126).

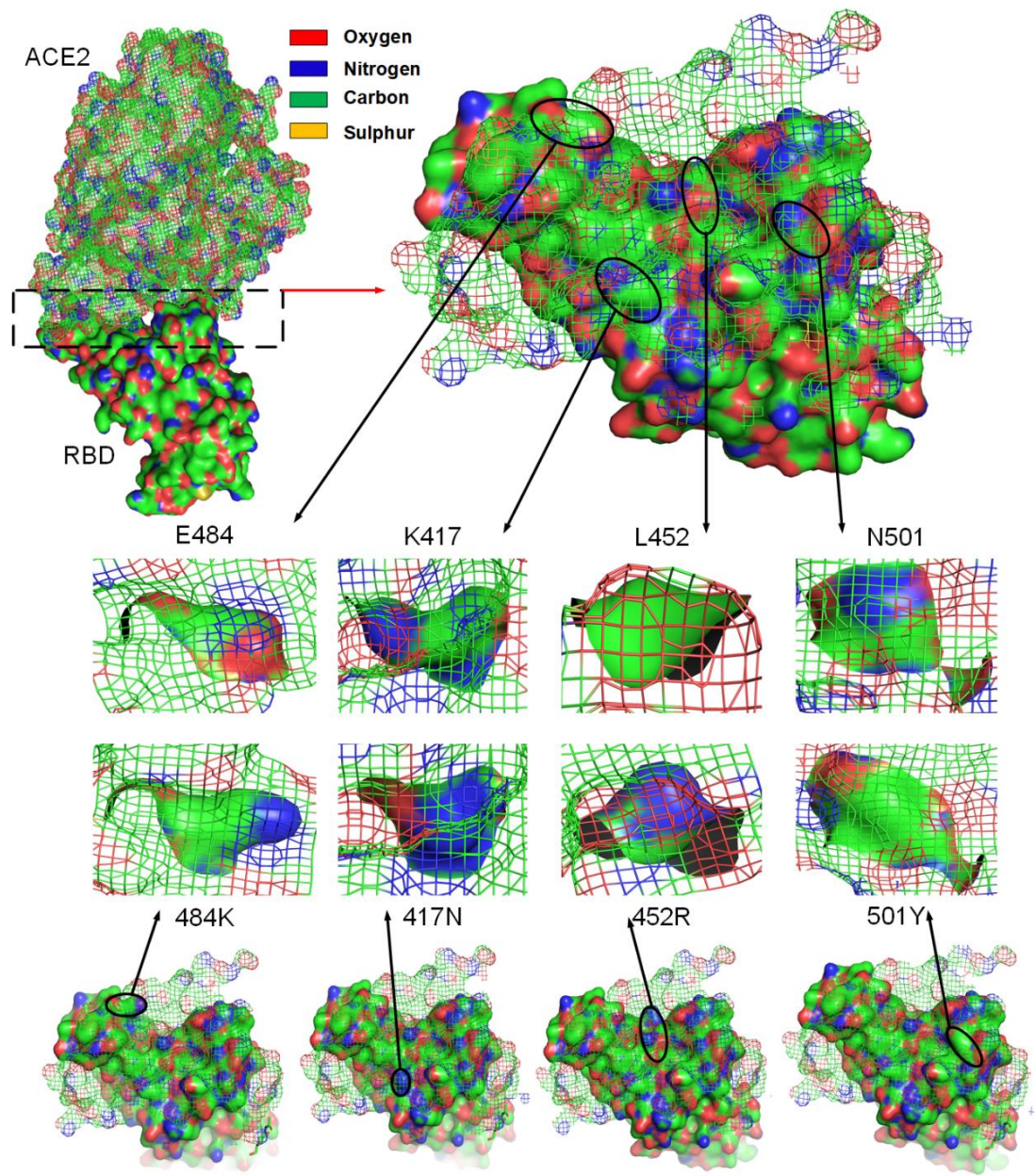

Fig.1 Comparison of distribution of hydrophobic (green areas) and hydrophilic (red and blue areas) surface areas on the RBD and ACE2 before and after the four mutations at the docking site (PDBID: 6LZG).

| Antibody | E484K | N501Y | K417N | L452R |
|---|---|---|---|---|
| ACE2 | 1.42 | 41.06 | -4.46 | 3.65 |
| CB6 | 23.58 | 9.22 | 0.02 | -6.06 |
| SR4 | 13.02 | -0.78 | 6.11 | -35.32 |
| CV30 | 7.38 | -3.27 | 1.17 | -3.80 |
| CC12.1 | 10.87 | -9.06 | -1.26 | -6.01 |
| CC12.3 | -0.07 | 10.65 | -9.79 | -5.50 |
| P2B-2F6 | 9.68 | 0.00 | 0.00 | -35.26 |
| COVA2-39 | 4.61 | -1.36 | -12.73 | 2.44 |
| MR17-K99Y | 4.87 | 6.68 | 3.15 | -19.73 |
| B38 | -9.76 | 0.50 | -9.55 | -11.01 |

Table 1 Sum of increment of hydrophobic attraction area and decrement of hydrophilic repulsion area among the variants, the ACE2 and the antibodies due to mutations (Å$^2$)

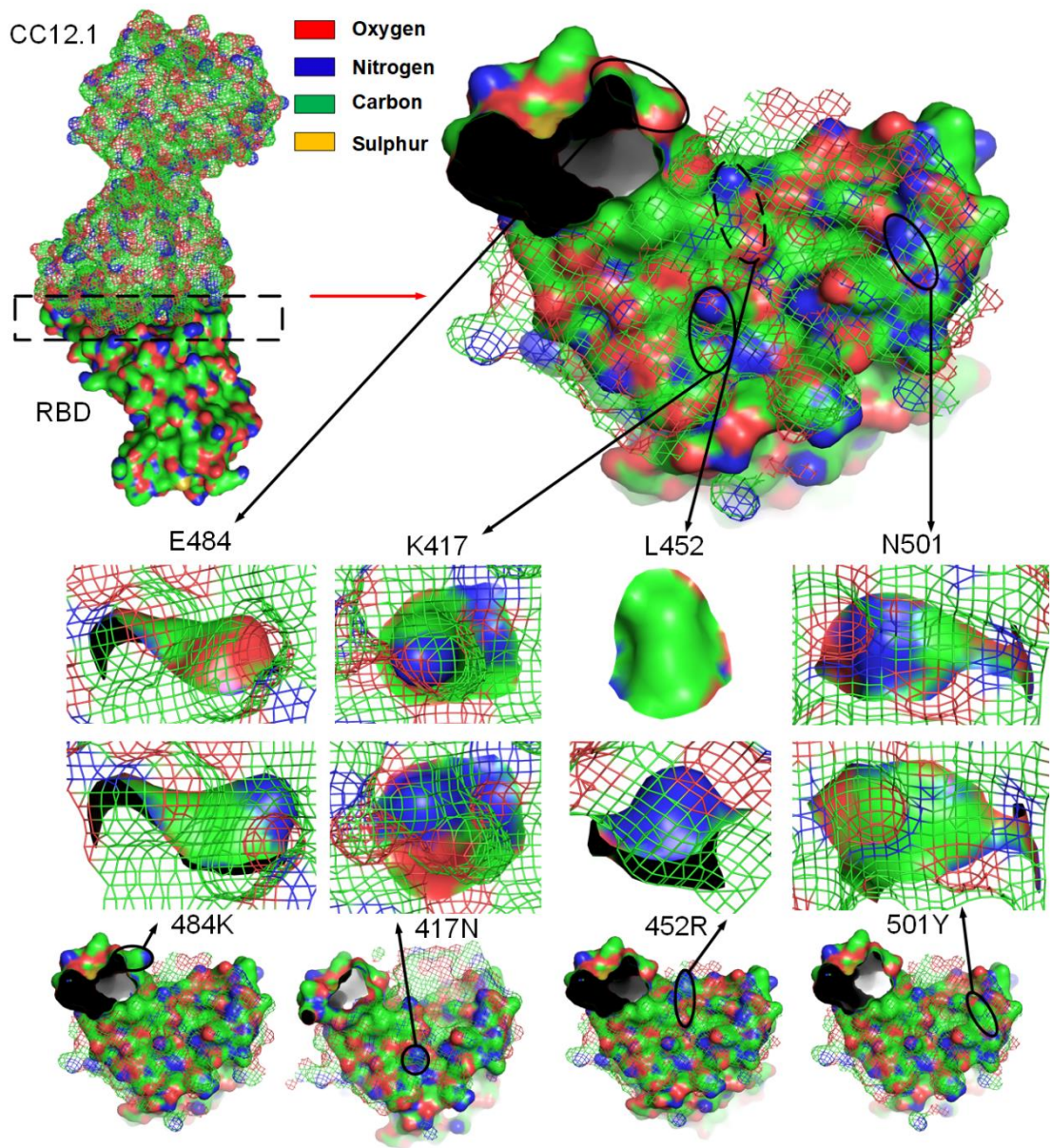

Fig.2 Comparison of distribution of hydrophobic (green areas) and hydrophilic (red and blue areas) surface areas on the RBD and antibody CC12.1 before and after the four mutations at the docking site (PDBID: 6XC2).

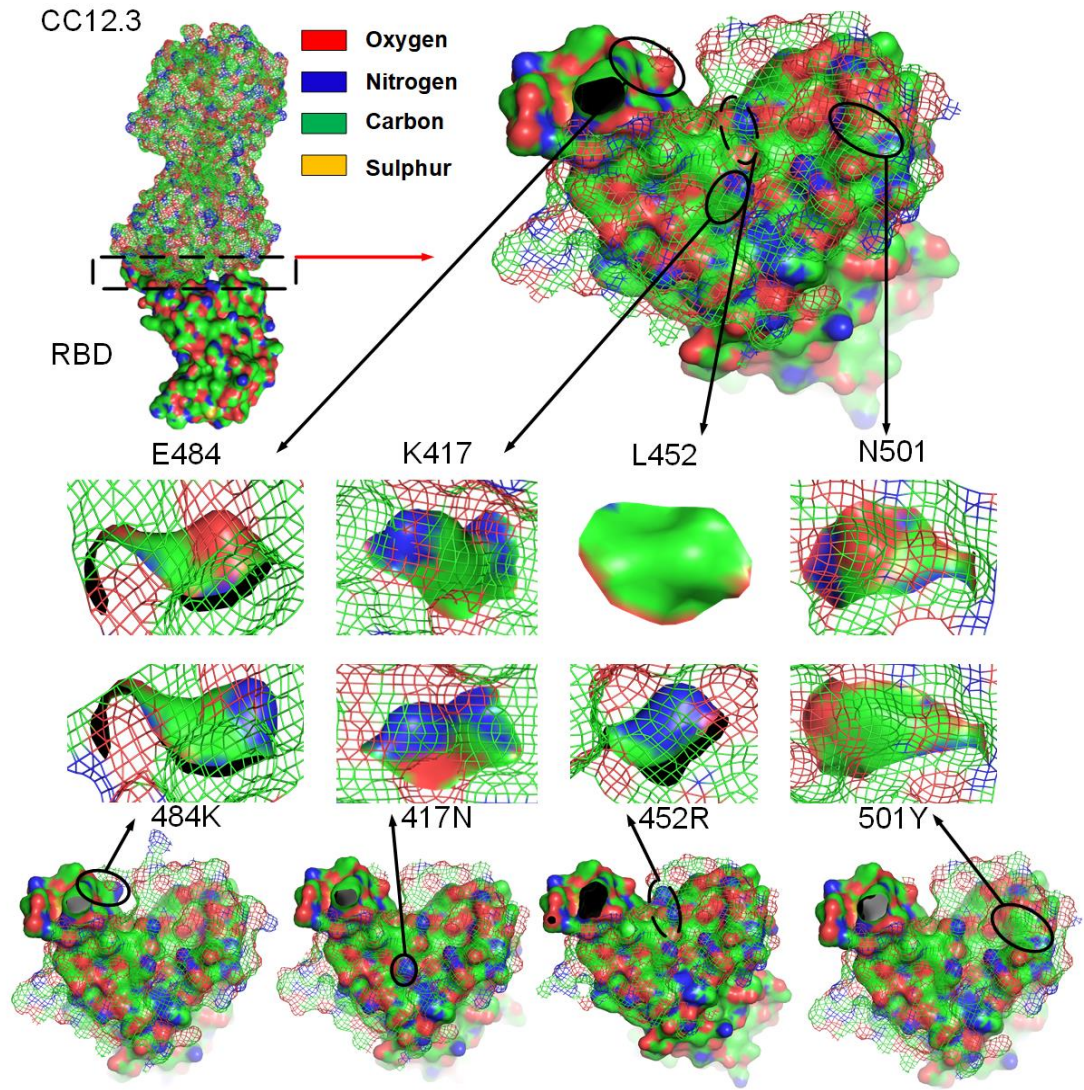

Fig.3 Comparison of distribution of hydrophobic (green areas) and hydrophilic (red and blue areas) surface areas on the RBD and antibody CC12.3 before and after the four mutations at the docking site (PDBID: 6XC4).

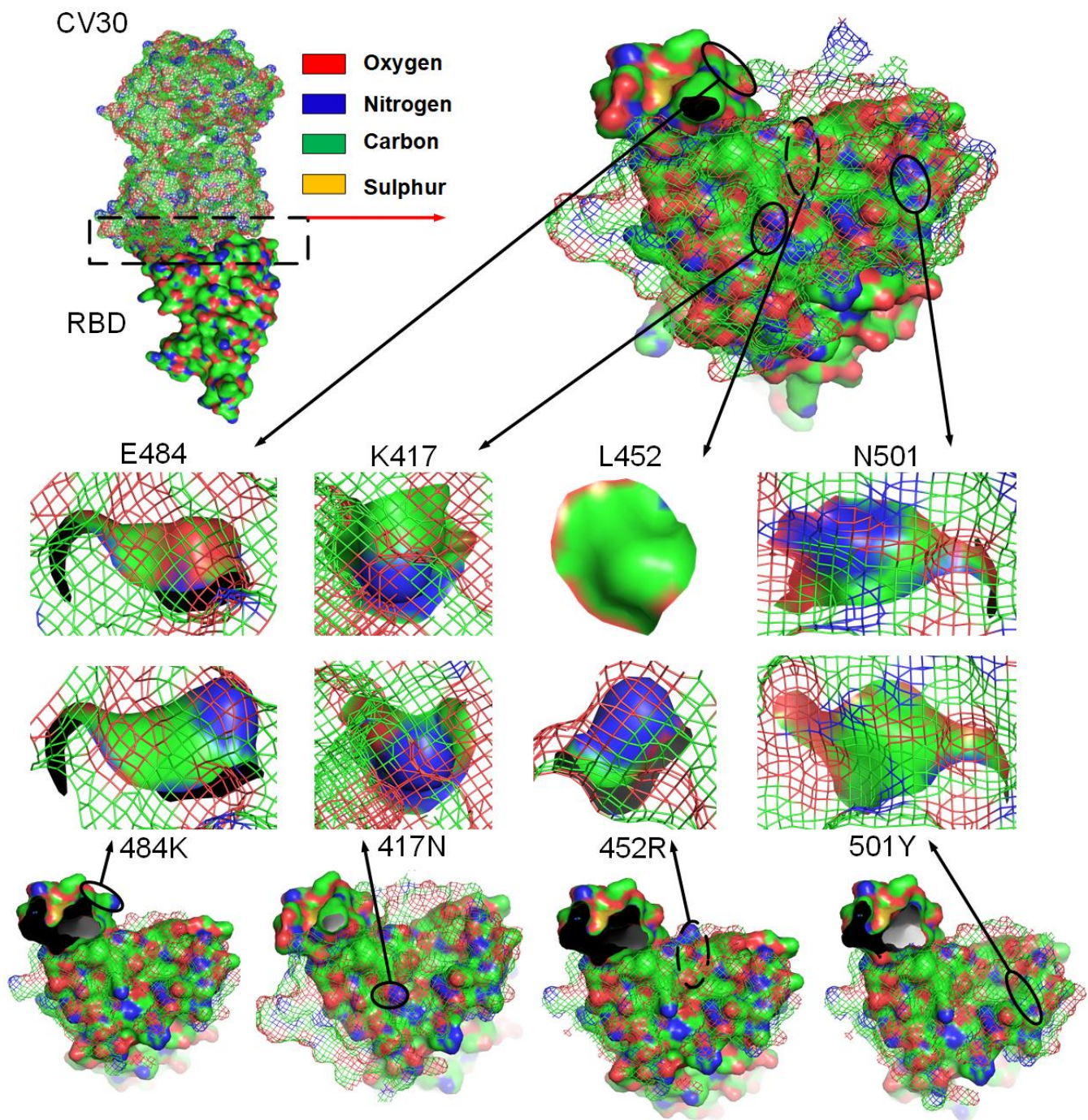

Fig.4 Comparison of distribution of hydrophobic (green areas) and hydrophilic (red and blue areas) surface areas on the RBD and antibody CV30 before and after the four mutations at the docking site (PDBID: 6XE1).

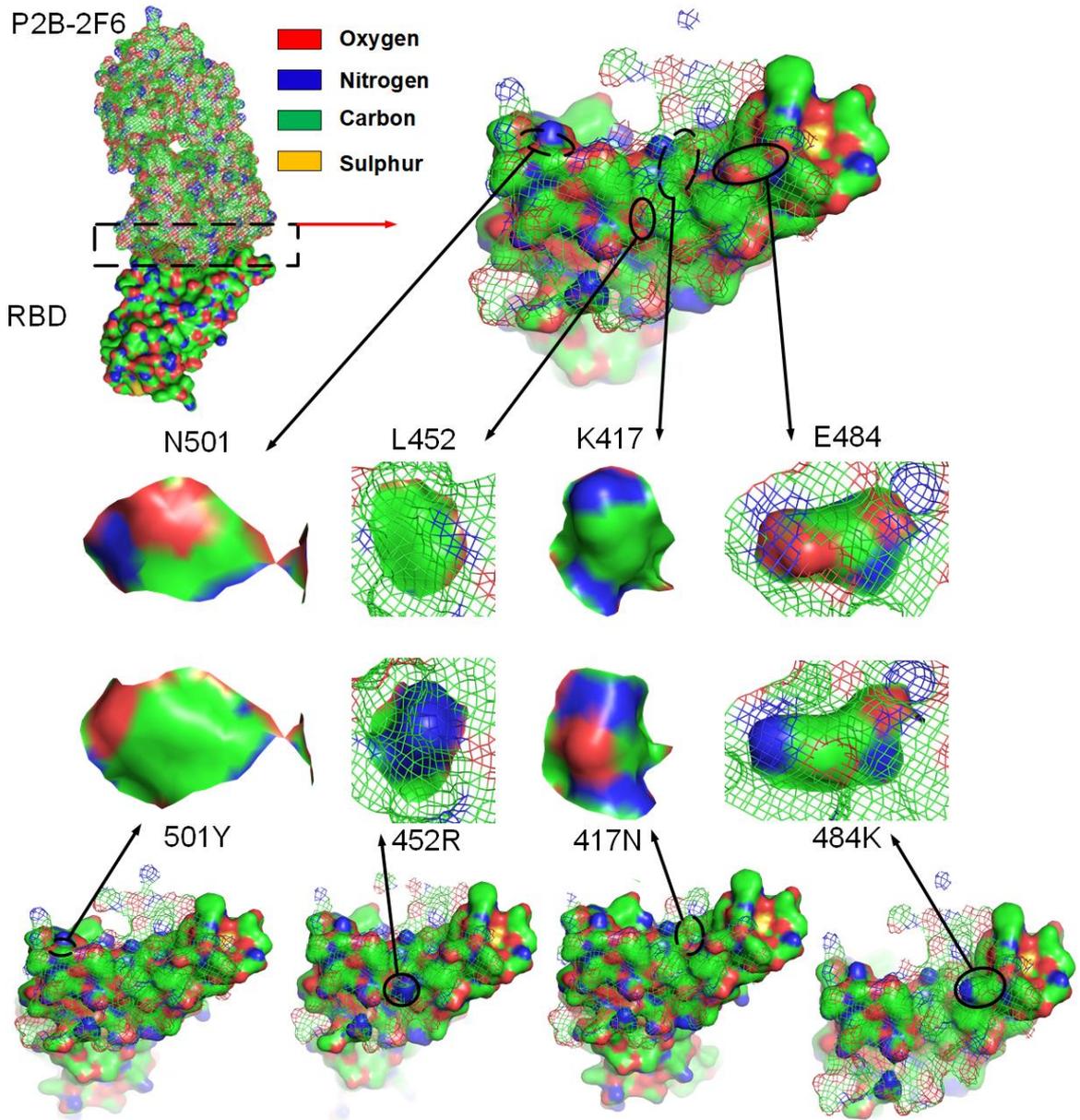

Fig.5 Comparison of distribution of hydrophobic (green areas) and hydrophilic (red and blue areas) surface areas on the RBD and antibody P2B-2F6 before and after the four mutations at the docking site (PDBID: 7BWJ).

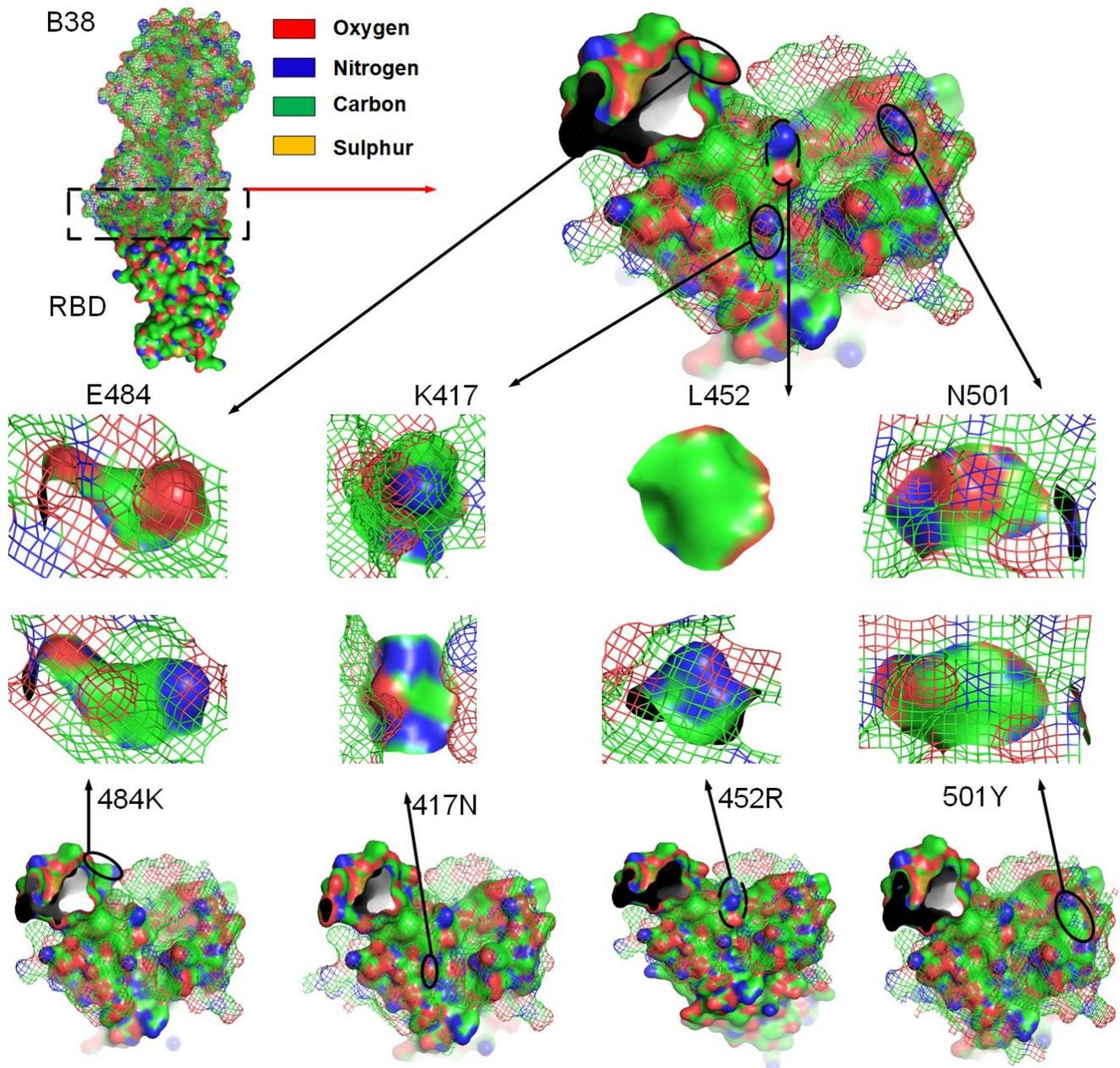

Fig.6 Comparison of distribution of hydrophobic (green areas) and hydrophilic (red and blue areas) surface areas on the RBD and antibody B38 before and after the four mutations at the docking site (PDBID: 7BZ5).

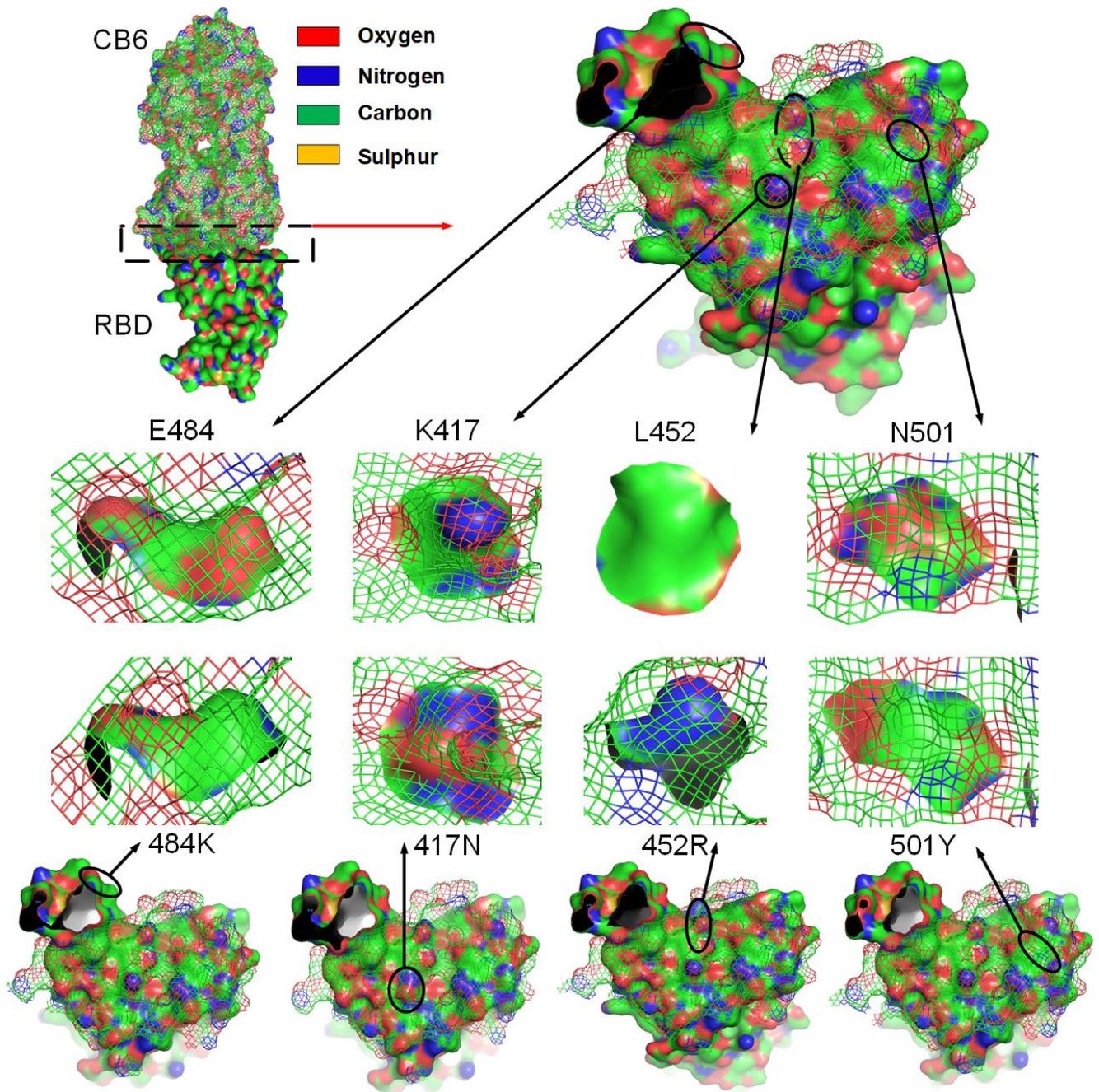

Fig.7 Comparison of distribution of hydrophobic (green areas) and hydrophilic (red and blue areas) surface areas on the RBD and antibody CB6 before and after the four mutations at the docking site (PDBID: 7C01).

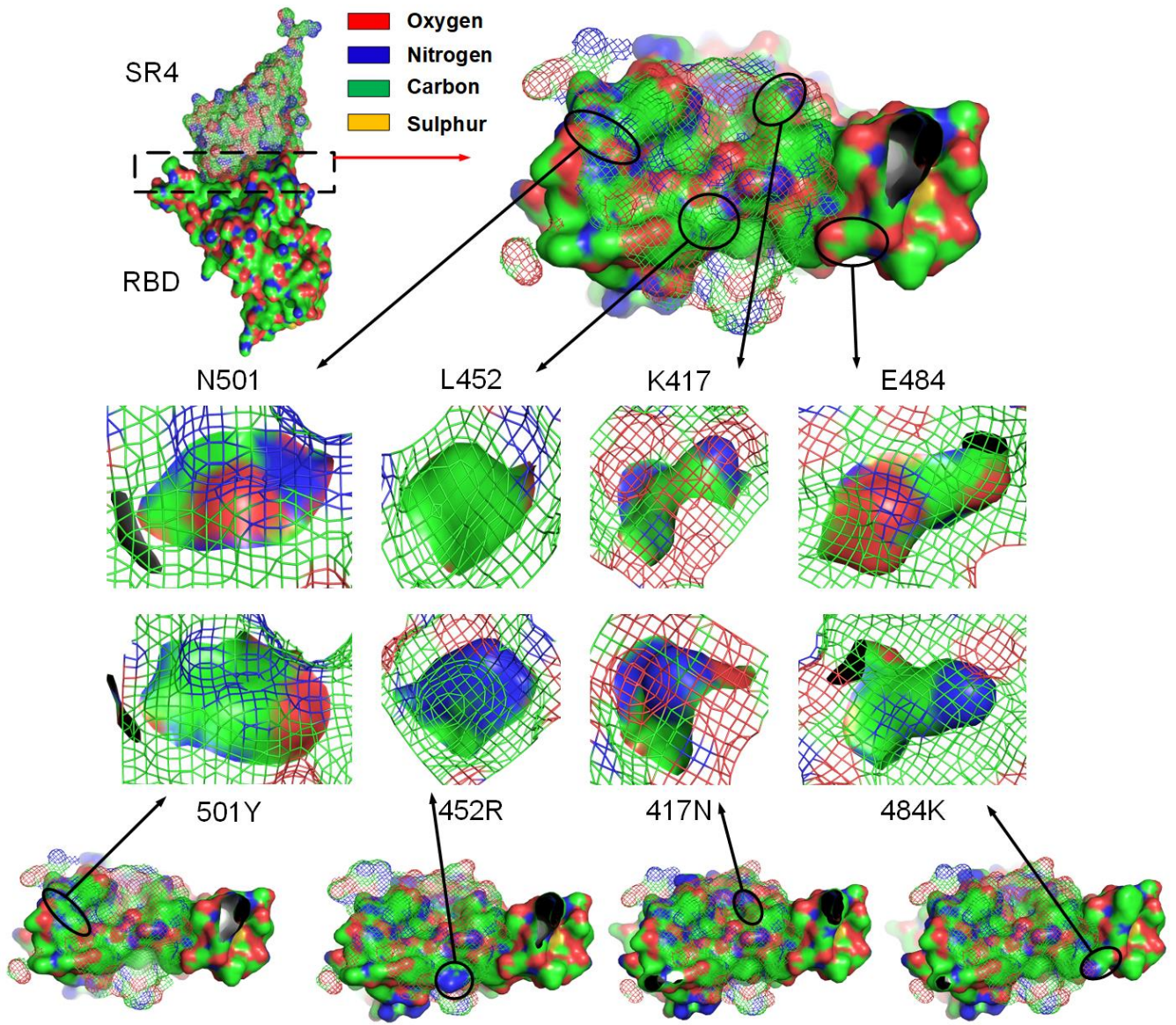

Fig.8 Comparison of distribution of hydrophobic (green areas) and hydrophilic (red and blue areas) surface areas on the RBD and antibody SR4 before and after the four mutations at the docking site (PDBID: 7C8V).

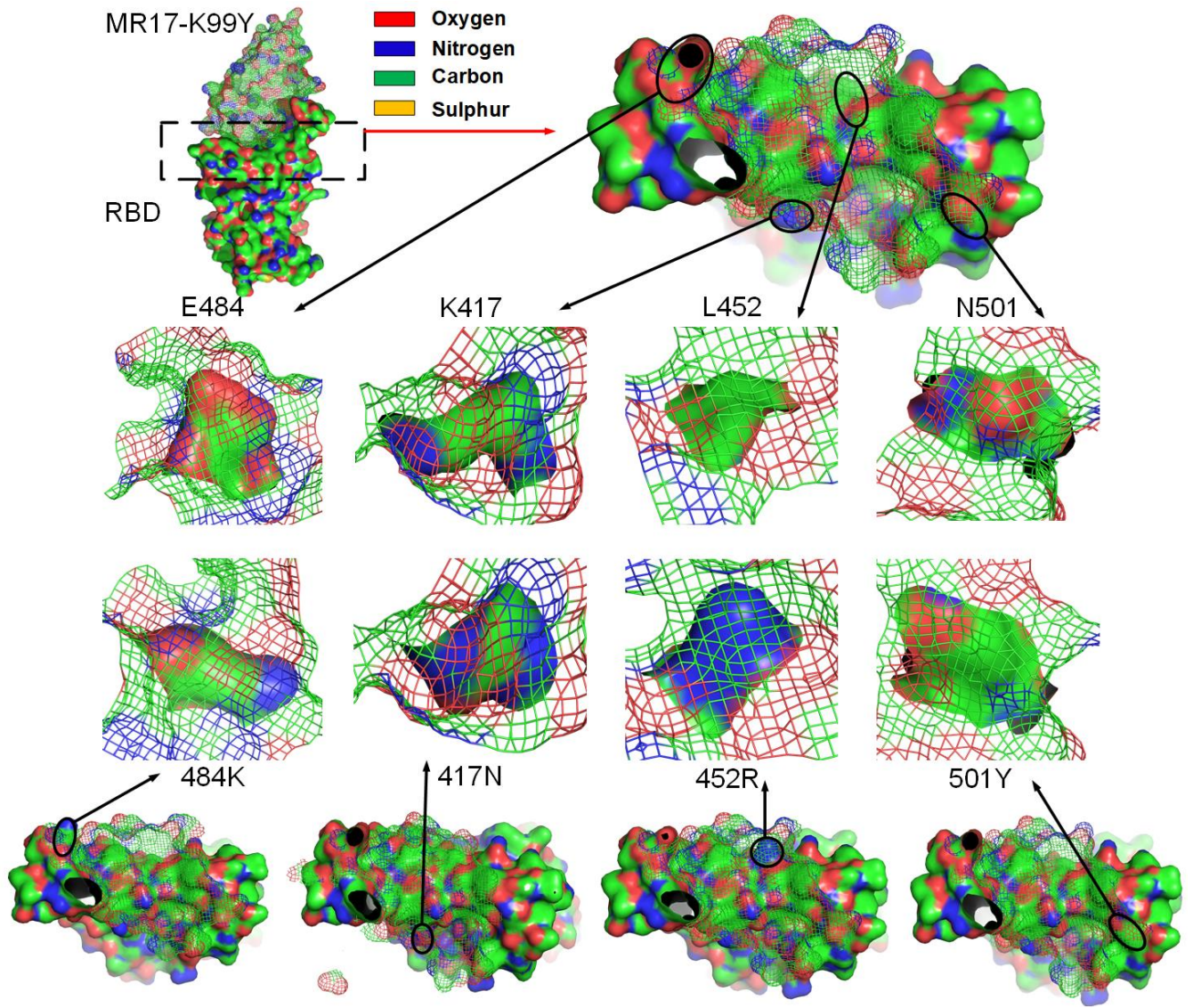

Fig.9 Comparison of distribution of hydrophobic (green areas) and hydrophilic (red and blue areas) surface areas on the RBD and antibody MR17-K99Y before and after the four mutations at the docking site (PDBID: 7CAN).

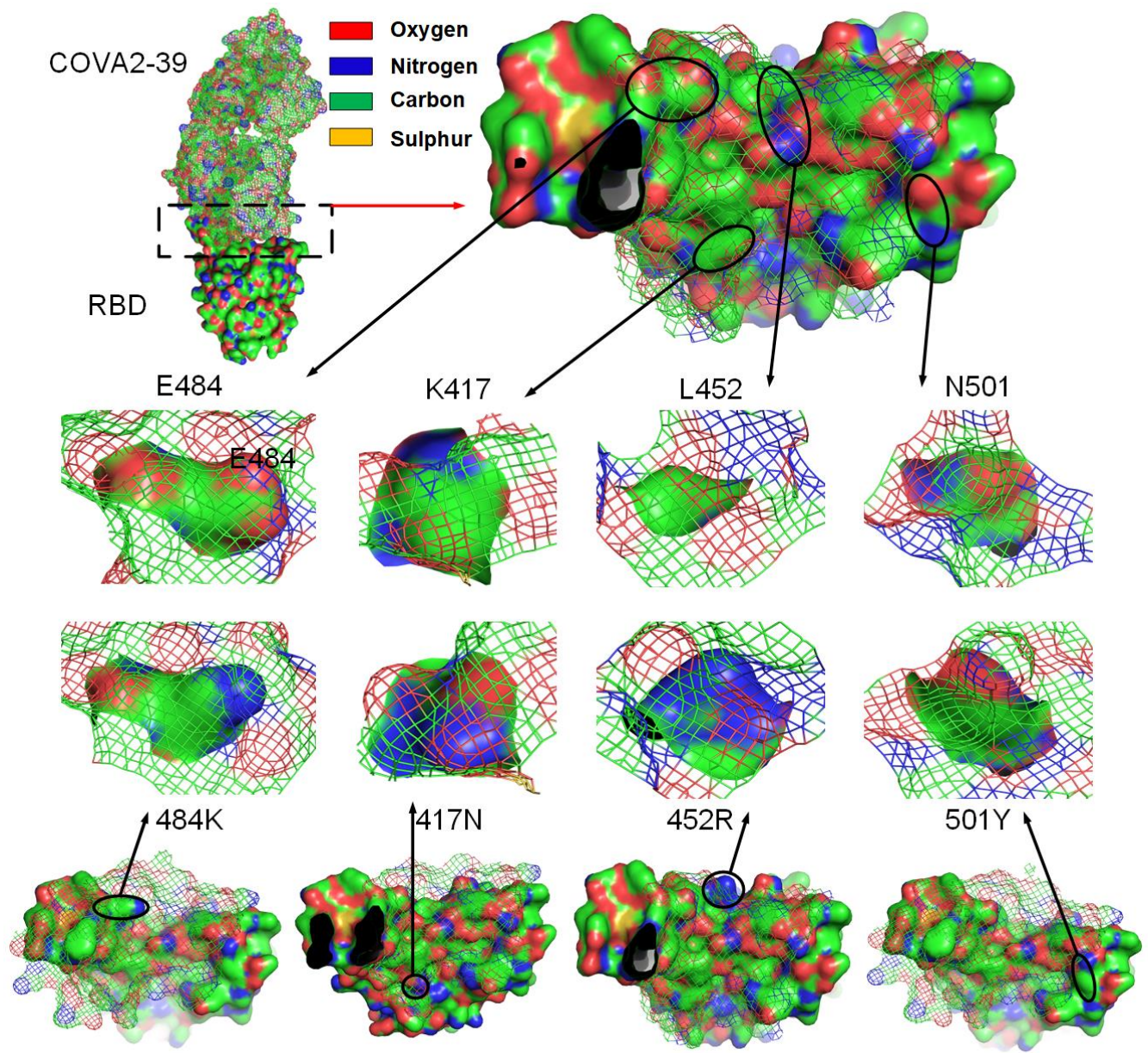

Fig.10 Comparison of distribution of hydrophobic (green areas) and hydrophilic (red and blue areas) surface areas on the RBD and antibody COVA2-39 before and after the four mutations at the docking site (PDBID: 7JMP).

**Additional Information**

The authors declare no competing financial interests.